\documentclass[5pt]{article}

\usepackage[letterpaper]{geometry}
\usepackage{spconf,amsmath,epsfig}
\usepackage{enumitem}
\usepackage{amsfonts}
\usepackage{booktabs}

\usepackage{fancyhdr}
\begin{document}\sloppy

\def\x{{\mathbf x}}
\def\L{{\cal L}}

\title{DeepQoE: A unified Framework for Learning to Predict Video QoE}
%

\name{Huaizheng Zhang$^1$, Han Hu$^1$, Guanyu Gao$^1$, Yonggang Wen$^1$ and Kyle Guan$^2$\thanks{Kyle Guan is the corresponding author.}}
\address{$^1$ Nanyang Technological University, Singapore\\
$^2$ Nokia Bell Labs, 791 Holmdel Road, Holmdel, NJ, USA\\
\{huaizhen001, hhu, ggao001, ygwen\}@ntu.edu.sg, kyle.guan@nokia.com}
%
%
%

\maketitle

\begin{abstract}
Motivated by the prowess of deep learning (DL) based techniques in prediction, generalization, and representation learning, we develop a novel framework called DeepQoE to predict video quality of experience (QoE). The end-to-end framework first uses a combination of DL techniques (e.g., word embeddings) to extract generalized features. Next, these features are combined and fed into a neural network for representation learning. Such representations serve as inputs for classification or regression tasks. Evaluating the performance of DeepQoE with two datasets, we show that for the small dataset, the accuracy of all shallow learning algorithm is improved by using the representation derived from DeepQoE. For the large dataset, our DeepQoE framework achieves significant performance improvement in comparison to the best baseline method (90.94\% vs. 82.84\%). Moreover, DeepQoE, also released as an open source tool, provides video QoE research much-needed flexibility in fitting different datasets, extracting generalized features, and learning representations.
\end{abstract}
\section{Introduction}
\label{sec:intro}
\vspace{-0.1in}
Video quality of experience (QoE), which assesses directly the perceptual quality of service (QoS) from the end users’ perspective, has become the de facto metric in guiding the design, deployment, and operation of video related services and applications \cite{balachandran2013developing, zhang2013qoe}. Notwithstanding its ever crucial role in services like video streaming, the measurement, modeling, and prediction of video QoE remain challenging tasks \cite{zhao2017qoe}. Video QoE depends on many often inter-related factors, from system parameters such as resolution and frame rate \cite{sidaty2014influence, huynh2008temporal} to demographic information such as gender and age \cite{huynh2008temporal}. These factors, often referred to as influence factors (IFs), fall into three categories: system IFs, context IFs, and human IFs. Though user experience is often considered as subjective and hard to quantify, human IFs will continue to be an essential part of QoE measurement and prediction \cite{zhang2013qoe}. For measuring QoE, two types of models, subjective test model and objective quality model, are often used \cite{chen2015qos}. Subjective test directly measures QoE by soliciting users’ evaluation under the controlled laboratory environment. Users are given a series of tested video sequences (original and processed ones) and then required to provide scores on the video quality. Objective quality models often use the results from subjective tests as ground truth to identify the objective QoS parameters that contribute to user perceptual quality and map these parameters to user QoE. Though these models are widely deployed, they have drawbacks. First, conducting subjective tests can be costly in terms of time and money. Second, these models often rely on hand-crafted features and data representations that are unique to the specific dataset, thus are difficult to be applied to other scenarios directly. As such, when applied to the new datasets, these models often do not generalize well. Third, though video QoE prediction can take the form of either classification or regression, both tasks often share significant similarities in features and processing methods. However, many models neglect these similarities and develop separate frameworks for feature engineering and training to perform classification or regression, often leading to inefficiency in the model developing and training process.

To bridge these gaps, we aim to design a QoE prediction framework with the following design guidelines and objectives. First, the use of dataset-specific representation and feature engineering should be minimized. Instead, by leveraging the potential of deep learning techniques and the availability of large datasets, both feature extraction and representation should be designed in an end-to-end learning based manner. Second, the framework should have the efficiency, configurability, and flexibility to perform multiple tasks and to facilitate transfer learning. To this end, we propose DeepQoE, an end-to-end and unified deep learning based framework that consists of three phases in tandem. First, we leverage deep learning based techniques (i.e. convolutional neural networks (CNNs)) to extract general features from different datasets or types of data. By applying these techniques, we can map data of different types and modalities all into the same high-dimensional feature space. Next, a deep neural network (DNN) is used to process different features to produce a representation. Finally, the existing or DeepQoE models can take directly the learned representations as input for classification or regression tasks. In a nutshell, the whole framework supplies a complete pipeline for feature extractions, representation learning, and QoE prediction, which is applicable for a variety of datasets.

\begin{figure}
  \includegraphics[width=\linewidth]{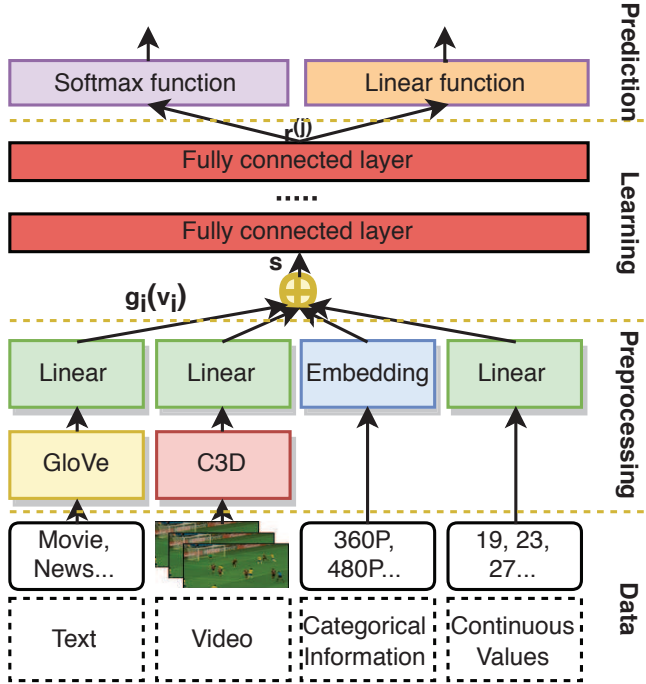}
  \centering
  \caption{DeepQoE framework. Input data, categorized into four types, is first processed with different learning based methods and then concatenated into one feature vector to learn a representation. The learned representation is used for classification or regression.}
  \label{fig:/DQ}
  \vspace{-0.25in}
\end{figure}

We compare DeepQoE to some shallow learning algorithms (e.g., decision tree) when solving classification problem. The results show that, in the small dataset, the performance of our framework is comparable to these non-deep-learning based algorithms. All machine learning algorithms perform better by using the representation derived from DeepQoE. When applying to a large dataset, DeepQoE achieves significant performance improvements compared to the strongest baselines (DeepQoE 90.94\% vs. SVM 82.84\%). In addition to the performance improvement, the proposed framework has the following advantages. First, to the best of our knowledge, DeepQoE is the first model to predict directly the QoE score, in the form of either classification or regression, using the same framework. Second, with the help a diverse set of deep learning techniques, DeepQoE provides powerful generalization and feature extractions, which enable effective transfer learning in video QoE research. To facilitate the QoE research, We also develop DeepQoE into an open source tool and release pre-trained DeepQoE models.

The rest of the paper is organized as follows. Section II introduces related works for deep learning and video QoE prediction. Section III provides a detailed description of the design of the DeepQoE framework. Section IV shows the details of experiments and presents experiments results. Section V concludes this paper and discusses future works.
\vspace{-0.12in}

\section{Related Works}
\label{sec:related}
\vspace{-0.1in}

\textbf{Deep Learning.}
Deep learning has emerged as a powerful set of frameworks and techniques that are widely applied to the research of computer vision, natural language processing, and speech recognition. In this work, we focus on three deep learning techniques: CNNs, word embeddings, DNNs. \textbf{CNNs} are the most powerful tool to process visual information and often provide much better results than traditional models in video analysis task \cite{karpathy2014large, tran2014c3d}. Among various CNN frameworks aimed to extract video features, DeepVideo \cite{karpathy2014large} first use CNN to extract features frame-by-frame and then fuse them to derive temporal correlations. 3-Dimensional Convolutional Neural Networks (C3D) \cite{tran2014c3d}, with an added dimension in the convolution filters, produces features that contain both frame and context information. \textbf{Word embeddings} map words into vectors, such that vectors of words that are similar in semantics are also close in distance. Two prevailing frameworks are word2vec \cite{mikolov2013distributed} and GloVe \cite{pennington2014glove}. Word2vec, a supervised learning based framework, uses a very large-scale dataset to learn word embeddings. In comparison, GloVe is an unsupervised learning based approach that uses co-occurrence statistics to produce word vectors. \textbf{DNNs}, also called deep forward networks, are often used function approximators. Recently, many new techniques, such as Dropout \cite{srivastava2014dropout}, are introduced to overcome the issues of over-fitting.

\textbf{Video QoE Prediction.}
The prevailing approaches for video QoE prediction in general fall into two categories: objective QoE monitoring and data-driven approaches. Objective QoE monitoring often considers a set of video QoE influence factors (IFs) and design schemes to fit them. \cite{malekmohamadi2012automatic} selects video content as the IF and propose a clustering algorithm to predict QoE. In \cite{song2016qoe}, a user-centric model is built to select the important external audiovisual factors and users' internal factors. Among data-driven approaches, some machine learning algorithms (e.g., linear regression \cite{dobrian2011understanding}, decision tree \cite{balachandran2013developing}) has been applied to predict QoE and to identify the important factors. Recently, deep learning techniques such as recurrent neural networks \cite{bampis2018recurrent} have been used to predict video QoE. Since most of these models or frameworks often rely on features unique to the particular dataset used, they may lack the capability to generalize. In addition, the models are designed to solve one task only: either classification or regression. To address these issues, we propose our DeepQoE framework, which can not only process data with a wide range of formats and modality and but also has the flexibility of performing both classification and regression tasks.
\vspace{-0.1in}

\section{Framework Design}
\label{sec: design}
\vspace{-0.1in}

The architecture of the proposed framework is illustrated in Fig.~\ref{fig:/DQ}. It has three phases that supply an end-to-end pipeline for predicting video QoE: feature preprocessing, representation learning, and QoE prediction.
\vspace{-0.15in}

\subsection{Feature preprocessing}
\vspace{-0.05in}

The goal of feature preprocessing phase is to map input data into initial feature vectors, which are to be fed into the representation learning phase. Since the training datasets could come from different sources, they pose the challenges for feature preprocessing in regard to the following aspects:
\begin{itemize}
\item Heterogeneity in data modality and type: some datasets only contain categorical information and numerical values; while other datasets include video sequences (or video features) as well as detailed text descriptions of video type and content.
\vspace{-0.1in}

\item Heterogeneity in representation approaches: even within a dataset, finding a general representation for different categorical information can be difficult. For example, while it is easy to encode user gender information with 0 and 1, it is less straightforward to encode the resolution information (e.g., 480P, 720P, and etc.) in an efficient manner. Moreover, categorical information such as video type can be represented as an index (integer), one-hot vector (vector of zeros and one), or an embedded vector (vector of continuous variables). It is not immediately obvious which representation will give rise to the best classification or regression performance.
\end{itemize}

To address these challenges, we categorize the input data into four types: text, video, categorical information (integer values), and continuous values. For each input type, we adopt a specific approach to extract the features. Specifically, we use GloVe (pre-trained on Wikipedia corpus), C3D (pretrained on Sport-1M dataset), embedding layer, and dense layer to extract the features for text, video, categorical information, and continuous values, as shown in Fig.~\ref{fig:/DQ}. Let $\mathbf{x_i}$ denote the input data of type $i$ and $\mathbf{v_i}$ denote the extracted feature vector, the prepossessing can be summarized as:
\begin{eqnarray}
\mathbf{v_i} = f_i( \mathbf{x_i}; \delta ) \in \mathbb{R}^{U},
\end{eqnarray}
where $f_i$ represents the feature extraction method for data type $i$ and $\delta$ represents the learned parameters.
\vspace{-0.1in}

\subsection{Learning representation}

In this phase, different feature vectors output by the preprocessing phase are firstly fused into a single feature vector. We use a simple concatenation to combine different feature vectors (we find other fusion approaches (such as 1D CNN \cite{bohez2017sensor}) can not offer noticeable performance improvement). Specifically, the fusion operation has the following mathematical form:
\begin{eqnarray}
\mathbf{s} = C ( g_1(\mathbf{v_1}), g_2(\mathbf{v_2}), ..., g_i(\mathbf{v_i})),
\end{eqnarray}
where $\mathbf{s}$ is the fused feature vector, $C$ represents concatenation operation, and $g_i$ is a general function associated with feature $\mathbf{v_1}$. The salient feature of this design is that $g_i(\mathbf{v_i})$ provide a general and flexible way of assigning different ``weights" for different features. Moreover, the choices of $g_i$ can be considered as a form of hyper-parameter tuning in the training process. This approach can help us not only to achieve better performance but also to identify the important contributing factors to QoE (by evaluating the $g_i(\mathbf{v_i})$ that achieve the best performance) during training and testing.

The fusion layer is followed by a few fully connected layers to continue the learning of a representation. The number of layers is another design parameter that can be adjusted depending on the size of the dataset. In particular, the representation out at layer $j$, $\mathbf{r}^{j}$, has the following form:
\begin{eqnarray}
\mathbf{r}^{(j)} = f (\mathbf{W}^{(j)} \mathbf{s}^{(j-1)}  + \mathbf{b}^{(j)}),
\label{drop}
\end{eqnarray}
where $\mathbf{s}^{(j-1)}$, $\mathbf{W}^{j}$, $\mathbf{b}^{j}$, and  $f$ represent the input, the weight, the bias, and the activation function of layer $j$. In addition, dropout technique \cite{srivastava2014dropout} is applied these hidden layers when training to prevent overfitting.
\vspace{-0.1in}

\begin{figure}
  \includegraphics[width=0.8\linewidth]{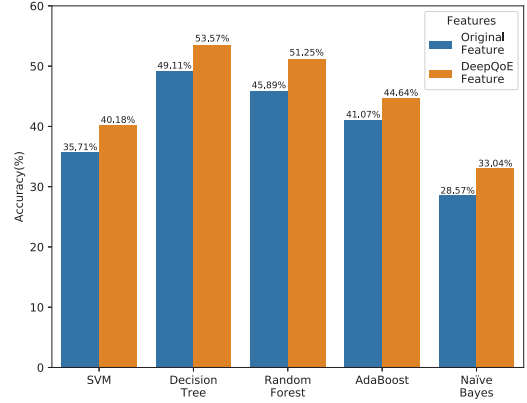}
  \centering
  \caption{Performance comparison between using original features and using representations derived from DeepQoE. Using DeepQoE representations, all models achieve better performance.}
  \label{fig:/whu_2}
  \vspace{-0.2in}
\end{figure}

\subsection{Predicting video QoE}

In this phase, the learned representation is fed into a NN or a DNN, which performs either classification or regression. Let $\mathbf{s}^{(l)}$ denote the representation vector output from representation learning phase and $\mathbf{Y}$ denote the ground-truth. Using a NN with only a single layer (specified by the weight matrix $\mathbf{W}^{l+1}$) as an example, we apply cross-entropy as loss function for classification:
\begin{eqnarray}
\mathcal{L} = - \mathbf{Y}\log{\mathcal{F}(\mathbf{W}^{l+1} \mathbf{s}^{(l)} + \mathbf{b}^{l+1})},
\label{cross}
\end{eqnarray}
where $\mathcal{F}$ is the \textit{softmax} activation function.
For regression, the loss function is:
\begin{eqnarray}
\mathcal{L} = \frac{1}{m} (\mathbf{Y} - \mathcal{F}(\mathbf{W}^{l+1} \mathbf{s}^{(l)} + \mathbf{b}^{l+1}))^2,
\label{linear}
\end{eqnarray}
where $\mathcal{F}$ is the \textit{linear} activation function and $m$ is the number of samples in a training batch.

\section{Results Presentation}

To evaluate the performance of the proposed DeepQoE framework, we design three experiments based on two datasets. For each experiment, the architecture of DeepQoE is adjusted according to the requirement of the experiment.
\vspace{-0.1in}

\subsection{Small text dataset}
The first and second experiments both use a small data set WHU-MVQoE2016 \cite{chenWHU}. There are four video types: movie, cartoon, news, and sports. For each type, there are two different video titles. Each video title is encoded with three resolutions: 720P, 480P, and 360P. In addition, each resolution is encoded with three different bitrates. In total there are 72 video clips in the dataset. A total of 16 subjects are asked to watch videos on the phone and rate them in the score of one (bad) to five (excellent). The dataset also includes the ages and genders of the end users. After post-processing, the datasets contains 1116 rating scores and 72 mean opinion scores (MOS).
\vspace{-0.15in}

\begin{figure}
  \includegraphics[width=0.8\linewidth]{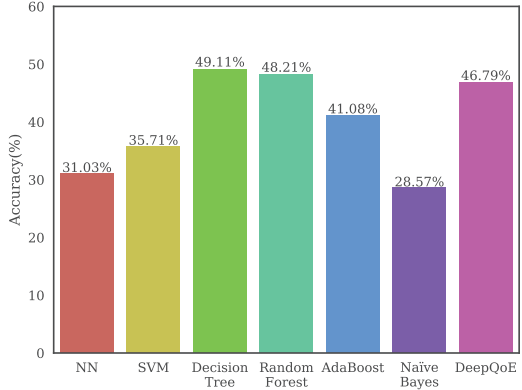}
  \centering
  \caption{Performance comparison of different methods on the small dataset. DeepQoE is comparable to the shallow learning algorithms by using same original features.}
  \label{fig:/whu_1}
  \vspace{-0.2in}
\end{figure}

\subsubsection{Classification}
The first experiment is conducted to predict users' voting scores,  which can be cast as a classification task of five classes. In the pre-processing module, we use pre-trained GloVe model to transform four video types to four 50-dimension vectors. For resolution, we use an embedding layer to map a resolution value to a vector of eight dimensions. For bitrate and user age, we normalize them to range {[}0, 1{]} and use the dense layer to get two vectors of one dimension; For user gender information, we use an embedding layer to get a vector of one dimension. Next, the 50-dimension vector of video type is reduced to 5-dimension. The representation learning phase concatenates these vectors into a single one and then feed it to two fully connected layers, with dropout technique applied to prevent overfitting. Finally, the output layer use softmax activation function and cross entropy loss function of training and prediction.

\begin{figure}
  \includegraphics[width=0.8\linewidth]{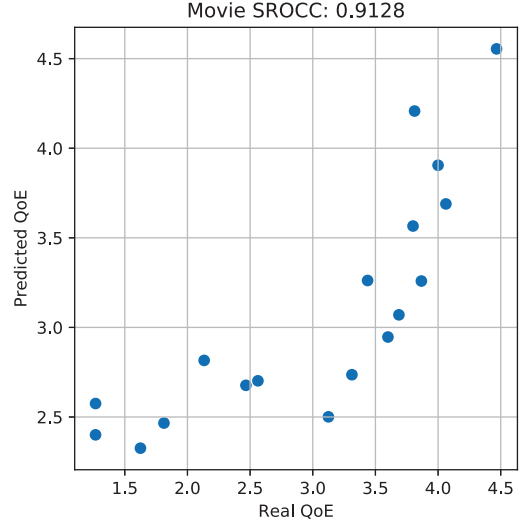}
  \centering
  \caption{Spearman Rank Order Correlation Coefficient (SROCC) of our DeepQoE trained by using news, cartoon and sport videos and tested on the movie videos.}
  \label{fig:/movie}
  \vspace{-0.2in}
\end{figure}

We compare the DeepQoE model with and other shallow learning models (such as SVM and random forest). In particular, we repeat the same experiment 10 times and use the average as our final result. Each time, the output of the last fully connected layer of the trained DeepQoE model is fed into one of the well-known machine learning models as features. The results show that all models that use representations provided by DeepQoE perform better than using original features (Fig.~\ref{fig:/whu_2}). By using original features directly, DeepQoE may not perform the best, but comparable to the shallow learning algorithms as demonstrated in Fig.~\ref{fig:/whu_1}.
\vspace{-0.1in}

\subsubsection{Regression}

\begin{table*}[]
\centering
\caption{Regression models comparison.}
\label{mse}
\begin{tabular}{cccccccccc}
\toprule[1pt]
Video\_1 & Video\_2 & Video\_3 & Video\_4 & Video\_5 & Video\_6 & Video\_7 & Video\_8 & \textbf{Baseline} & DeepQoE \\ \midrule[0.5pt]
0.104 & 0.145 & 0.107 & 0.076 & 0.129 & 0.076 & 0.141 & 0.229 & \textbf{0.126} & 0.298 \\ \bottomrule[1pt]
\end{tabular}
\vspace{-0.2cm}
\end{table*}

In the second experiment, we predict the MOS of the videos. Since this is cast as a regression problem, the ground truth is replaced by average scores and data related to user gender and age is not used.

We use the simple regression model from \cite{zhang2013qoe} as the baseline. The model essentially performs single variable regression --- it predicts MOS based only on the bitrate of a video title. We first apply this regression model to each of the eight video titles in our dataset and generate eight mean square error (MSE) values. We then take the average of these eight MSEs, which is 0.126. To evaluate the performance of DeepQoE, we use all the video related information in our dataset (video type, resolution, and bitrate) and train the model only once in an end-to-end fashion. The MSE of DeepQoE regression is 0.298, as shown in Table~\ref{mse}.

\begin{table}[]
\centering
\caption{DeepQoE regression results. We use three kinds of videos as the training set and then use the remaining one as the test set to verify that our approach can get a fair prediction result.}
\label{remain-one}
\begin{tabular}{cccc}
\toprule[1pt]
Movie & Cartoon & Sport & News \\ \midrule[0.5pt]
0.4119   & 0.5069   & 0.6771   & 1.2679   \\ \bottomrule[1pt]
\end{tabular}
\vspace{-0.2cm}
\end{table}

The larger MSE of our model can be initially attributed to the fact that deep learning based models often work better with a very large dataset, as evident by the results to be presented in the next section. Improving the regression performance (with the help of new techniques and more data) remains the focus of our ongoing research. However, we note that our model has two distinctive advantages. First, the regression performed by the baseline model is on a per-video basis, thus the regression coefficients obtained for one video cannot be directly applied to another one. In contrast, our DeepQoE model, trained on one dataset, includes all the general features sufficient to predict MOS of videos of various type, resolution, and bitrate (Table.~\ref{remain-one}). Second, for a new video sequence of given bitrate, baseline model cannot directly predict the MOS of this video, due to the unavailability of data points of different bitrates (for obtaining the regression coefficients). With the help of generalized feature extractions, our model can find the correlation between the new video and trained ones, and thus is able to predict the MOS (Figure.~\ref{fig:/movie}).
\vspace{-0.1in}

\subsection{Large video dataset}

\begin{table*}[]
\centering
\caption{Training time comparison.}
\label{time_compare}
\begin{tabular}{cccccc}
\toprule[1pt]
SVM      & Decision Tree & Random Forest & AdaBoost & Na\"{\i}ve Bayes & \textbf{DeepQoE} \\ \midrule[0.5pt]
3344.64s & 82.91s        & 33.56s        & 700.03s  & 233.54s     & \textbf{504.41s} \\ \bottomrule[1pt]
\end{tabular}
\vspace{-0.2cm}
\end{table*}

\begin{figure}
  \includegraphics[width=0.8\linewidth]{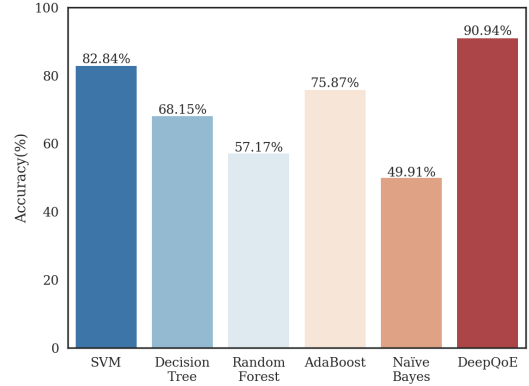}
  \centering
  \caption{Performance comparison on the large dataset. DeepQoE shows the best performance when comparing to the other shallow learning algorithms.}
  \label{fig:/jaykuo_1}
  \vspace{-0.2in}
\end{figure}

For the third experiment, we use a large dataset from \cite{wang2017videoset}, which has 220 different video titles. Each video title is five seconds in duration and has four different resolutions. For each resolution, a video title is encoded into 52 video clips with different quantization parameters (QP). In total, there are 45760 different video clips in the data set. A total of 800 subjects participate in this test. After post-processing, three just-noticeable-difference (JND) \cite{fischer2014serial} points are derived for each video clip. A JND point is a statistical quantity that accounts for maximum difference unnoticeable by a user. Using the notion of JND to measure the quality of coded images and videos was recently proposed \cite{xue2012mobile,wu2013just}.  As such, we use JND as QoE metric in our experiment. Specific to our experiment and dataset \cite{wang2017videoset}, the three JND points represent the three QP parameters at which noticeably degrade of video quality is observed. That is, any QP values smaller than (before the occurrence) of the first JND point is considered excellent; QP values that are in between the first and second JND points is considered as good; QP values that are in between the second and third JND points is considered as fair; the QP values larger than the third JND points are considered as bad. Thus we cast the QoE predication as a classification problem with four QoE classes: excellent, good, fair, and bad. Similar to the previous two experiment, we extract generalized features first and then make a prediction using softmax activation function. Since this dataset includes the video clips in addition to categorical data, we take full advantage of this by using 3D CNN to extract video content feature. The result shows that DeepQoE model can effectively capture the JND information and thus provide much better classification performance — the 90.94\% accuracy is higher than all of those generated by shallow learning algorithms as shown in Fig~\ref{fig:/jaykuo_1}. In comparison to the results generated by using small dataset, the proposed deep QoE method performs the best in the large dataset. Moreover, training of the proposed model takes only about one-sixth of time use for training with SVM, which provides the best performance in non-deep-learning based algorithms (Table.~\ref{time_compare}).
\vspace{-0.1in}

\section{Conclusion and Future Works}

Accurate and efficient QoE prediction provides important guidance to the deployment and operation of video services and applications. To address two main drawbacks of the current prediction models, namely over-reliance on dataset-specific feature engineering and lack of the configurability for transfer learning, we propose DeepQoE, a deep learning based framework capable of feature extraction, representation learning, and QoE prediction. Our results show that the learned representation via DeepQoE can improve the prediction accuracy of shallow learning models for a small dataset. When applied to a larger dataset, our framework is shown to achieve the best performance in comparison to other start-of-art algorithms. For our future research, we plan to continue stress-test and improve the performance of DeepQoE, as larger datasets of subjective QoE test become available. We also plan to extend DeepQoE to real-time to QoE prediction. In particular, we will evaluate if techniques such as attention mechanism \cite{bahdanau2014neural} can help to improve the performance in continuous-time scenarios.
\vspace{-0.1in}

\bibliographystyle{IEEEbib}
\bibliography{camera-ready_icme2018template}

\begin{thebibliography}{10}

\bibitem{balachandran2013developing}
Athula Balachandran and et~al.,
\newblock ``Developing a predictive model of quality of experience for internet
  video,''
\newblock in {\em ACM SIGCOMM Computer Communication Review}. ACM, 2013,
  vol.~43, pp. 339--350.

\bibitem{zhang2013qoe}
Weiwen Zhang and et~al.,
\newblock ``Qoe-driven cache management for http adaptive bit rate streaming
  over wireless networks,''
\newblock {\em IEEE Transactions on Multimedia}, vol. 15, no. 6, pp.
  1431--1445, 2013.

\bibitem{zhao2017qoe}
Tiesong Zhao and et~al.,
\newblock ``Qoe in video transmission: A user experience-driven strategy,''
\newblock {\em Commun. Surveys Tuts}, vol. 19, no. 1, pp. 285--302, 2017.

\bibitem{sidaty2014influence}
Naty~Ould Sidaty and et~al.,
\newblock ``Influence of video resolution, viewing device and audio quality on
  perceived multimedia quality for steaming applications,''
\newblock in {\em EUVIP}. IEEE, 2014, pp. 1--6.

\bibitem{huynh2008temporal}
Quan Huynh-Thu and et~al.,
\newblock ``Temporal aspect of perceived quality in mobile video
  broadcasting,''
\newblock {\em IEEE Trans. Broadcast}, vol. 54, no. 3, pp. 641--651, 2008.

\bibitem{chen2015qos}
Yanjiao Chen and et~al.,
\newblock ``From qos to qoe: A tutorial on video quality assessment,''
\newblock {\em Commun. Surveys Tuts}, vol. 17, no. 2, pp. 1126--1165, 2015.

\bibitem{karpathy2014large}
Andrej Karpathy and et~al.,
\newblock ``Large-scale video classification with convolutional neural
  networks,''
\newblock in {\em CVPR}, 2014, pp. 1725--1732.

\bibitem{tran2014c3d}
Du~Tran and et~al.,
\newblock ``C3d: generic features for video analysis,''
\newblock {\em CoRR, abs/1412.0767}, vol. 2, no. 7, pp. 8, 2014.

\bibitem{mikolov2013distributed}
Tomas Mikolov and et~al.,
\newblock ``Distributed representations of words and phrases and their
  compositionality,''
\newblock in {\em NIPS}, 2013, pp. 3111--3119.

\bibitem{pennington2014glove}
Jeffrey Pennington and et~al.,
\newblock ``Glove: Global vectors for word representation,''
\newblock in {\em EMNLP}, 2014, pp. 1532--1543.

\bibitem{srivastava2014dropout}
Nitish Srivastava and et~al.,
\newblock ``Dropout: a simple way to prevent neural networks from
  overfitting.,''
\newblock {\em Journal of machine learning research}, vol. 15, no. 1, pp.
  1929--1958, 2014.

\bibitem{malekmohamadi2012automatic}
Hossein Malekmohamadi and et~al.,
\newblock ``Automatic qoe prediction in stereoscopic videos,''
\newblock in {\em ICMEW}. IEEE, 2012, pp. 581--586.

\bibitem{song2016qoe}
Jiarun Song and et~al.,
\newblock ``Qoe evaluation of multimedia services based on audiovisual quality
  and user interest,''
\newblock {\em IEEE Trans. Multimedia}, vol. 18, no. 3, pp. 444--457, 2016.

\bibitem{dobrian2011understanding}
Florin Dobrian and et~al.,
\newblock ``Understanding the impact of video quality on user engagement,''
\newblock in {\em ACM SIGCOMM Computer Communication Review}. ACM, 2011,
  vol.~41, pp. 362--373.

\bibitem{bampis2018recurrent}
Christos~G Bampis and et~al.,
\newblock ``Recurrent and dynamic models for predicting streaming video quality
  of experience,''
\newblock {\em IEEE Transactions on Image Processing}, 2018.

\bibitem{bohez2017sensor}
Steven Bohez and et~al.,
\newblock ``Sensor fusion for robot control through deep reinforcement
  learning,''
\newblock {\em arXiv preprint arXiv:1703.04550}, 2017.

\bibitem{chenWHU}
Yingxue Zhang and et~al.,
\newblock ``Whu-mvqoe2016: A quality of experience dataset for mobile video
  research,''
\newblock Dec. 2016.

\bibitem{wang2017videoset}
Haiqiang Wang and et~al.,
\newblock ``Videoset: A large-scale compressed video quality dataset based on
  jnd measurement,''
\newblock {\em J Vis Commun Image Represent}, vol. 46, pp. 292--302, 2017.

\bibitem{fischer2014serial}
Jason Fischer and et~al.,
\newblock ``Serial dependence in visual perception,''
\newblock {\em Nature neuroscience}, vol. 17, no. 5, pp. 738--743, 2014.

\bibitem{xue2012mobile}
Jingteng Xue and et~al.,
\newblock ``Mobile jnd: Environment adapted perceptual model and mobile video
  quality enhancement,''
\newblock in {\em ACM MMSys}. ACM, 2012, pp. 173--183.

\bibitem{wu2013just}
Jinjian Wu and et~al.,
\newblock ``Just noticeable difference estimation for images with free-energy
  principle,''
\newblock {\em IEEE Trans. Multimedia}, vol. 15, no. 7, pp. 1705--1710, 2013.

\bibitem{bahdanau2014neural}
Dzmitry Bahdanau and et~al.,
\newblock ``Neural machine translation by jointly learning to align and
  translate,''
\newblock {\em arXiv preprint arXiv:1409.0473}, 2014.

\end{thebibliography}

\end{document}